\documentclass[aps,prl,twocolumn,groupedaddress,showpacs]{revtex4-1}

\usepackage{graphicx}
\usepackage{amsmath}
\usepackage{amssymb}

\begin{document}


\title{Shocks Generate Crossover Behaviour in Lattice Avalanches}


\author{James Burridge}
\email[]{james.burridge@port.ac.uk}
\affiliation{Department of Mathematics, University of Portsmouth, PO1 3HF, United Kingdom. }


\date{\today}

\begin{abstract}
A spatial avalanche model is introduced, in which avalanches increase stability in the regions where they occur. Instability is driven globally by a driving process that contains shocks. The system is typically subcritical, but the shocks occasionally lift it into a near or super critical state from which it rapidly retreats due to large avalanches. These shocks leave behind a signature -- a distinct  power--law crossover in the avalanche size distribution. The model is inspired by landslide field data, but the principles may be applied to any system that experiences stabilizing failures, possesses a critical point, and is subject to an ongoing process of destabilization which includes occasional dramatic destabilizing events.
\end{abstract}


\maketitle


\emph{Introduction.}-- In systems where failures can propagate, the final extent of the failure, however it is measured, often follows a power--law distribution. Such statistical behaviour has, for example, been observed in landslides \cite{Herg03,Pieg06}, earthquakes \cite{Pisa04,Olam92} electrical network failures \cite{Car02}, wildfires \cite{Rico01,Dros92}, and disease outbreaks \cite{Rhod96,Ben04}.  Reductionist avalanche models \cite{Har64,Herg02} suggest that power--law distributions appear when the ease with which failures propagate reaches a critical level toward which many such systems self--organise \cite{Dros92,Herg02,Olam92,Herg12}. When these systems are sub--critical, the power--law region is cut off, typically by an exponentially decaying probability density.

\begin{figure}
\includegraphics[width=8.6 cm]{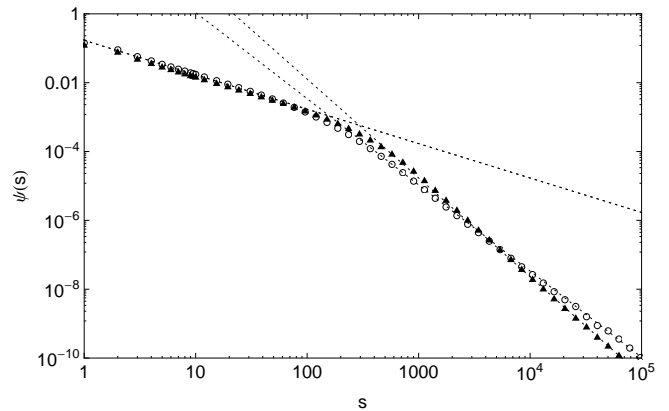}
\caption{\label{COpdf} Simulated steady state probability mass functions $\psi(s)$ for avalanche size in a system of size $W \times L = 2000 \times 3000$ driven by a compound Poisson process (triangles) with parameters $\nu=0.25$, $\lambda=10$ and $J_{max}=0.05$, and by a Gamma process (open circles) with parameters $\alpha=5$ and $\beta=10$, with $\epsilon =0.01$ in both cases. To reduce statistical noise, $\psi(s)$ has been estimated by averaging over intervals of the form $[\lfloor a^k \rfloor,\lfloor a^{k+1} \rfloor]$ where $a=1.25$. The dotted lines show, in the Poisson case, the pure power law approximations $\psi(s) \approx 0.18 \times s^{-1}$ valid for $s<230$ and $\psi(s) \approx 500 \times s^{-\tau}$ for $s>230$ where $\tau=2.9$ from equation (\ref{tau}). In the Gamma case we show the pure power law $\psi(s) \approx 3\times 10^2 \times s^{-\tau}$ valid for $s>140$ where $\tau = 2.5$ }
\end{figure}

In this work we investigate the phenomenon of power--law crossover. Here the failure size distribution, rather than having an exponential tail, is characterised by two different power--law exponents and the switch from one to the other occurs at a well defined size (see Figure \ref{COpdf}). Our investigation was inspired by the appearance of landslide inventory data \cite{Eeck07,Brar03,Whit13} showing that cumulative records of landslide areas can exhibit this phenomenon.

Crossover behaviour has been observed previously in the size distribution of fibre failure avalanches in fibre bundles, when the bundle is close to complete breakdown \cite{Prad06}.  In common with the fibre bundle case, the crossover in our model arises when the system is close to criticality. In contrast, failures drive our system away from criticality by locally reducing susceptibility to further failures. Our system is driven toward criticality by a global destabilization process, which may be thought of as performing the role of energy or particle addition in self--organising models \cite{Bak88,Olam92,Stan96,Pieg06}. The crucial ingredient in this destabilization process, which is responsible for the crossover, is the presence of jumps in instability, or ``shocks''. Without these the system would simply stabilize in a near critical state, producing the standard power--law size distribution, with exponential cutoff. The author's recent study of a non--spatial failure process driven by Brownian motion \cite{Burr13} laid some of the principles we use here.

Our model was inspired by landslide data, but there is experimental evidence of crossover behaviour in the distribution of wildfire areas \cite{Rico01} and the seismic moments of earthquakes \cite{Sorn96,Pisa04}. Data on the distribution of the sizes of measles outbreaks \cite{Rhod96,Jans03} is also suggestive of crossover. The shock--crossover relationship that we demonstrate could be present in any system that experiences stabilizing failures, possesses a critical point, and is subject to rapid destabilization events. Each of the physical systems just mentioned exhibits critical scaling and experiences catastrophes that reduce risk. In the case of wildfires, rapid increases in susceptibility could be caused by spells of particularly hot and dry weather. In  earthquakes, a jump in instability would correspond to a rapid increase in shear forces. In the case of disease outbreaks, a new disease strain could raise the disease transmission rate close to or above the critical epidemic threshold \cite{Ben04}. We therefore suggest that the principles of our model may have broad applicability.

\begin{figure}
\includegraphics[width=8.6 cm]{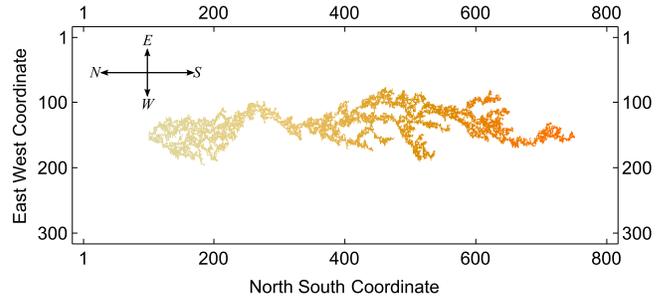}
\caption{\label{ava} An example of an avalanche in a $300 \times 800$ lattice when $p_{ij}=0.54$ for all sites. Later generations of the avalanche are more darkly shaded.}
\end{figure}

\emph{Avalanche construction.} -- We generate avalanches by a generalization of the classical branching process \cite{Har64,Stan96} to a rectangular $W \times L$ lattice with periodic boundary conditions where $W$ and $L$ are referred to as the \emph{east--west} and \emph{north--south} dimensions of the system (see Figure \ref{ava}). We suppose that avalanches propagate under the influence of a force (gravity in the case of landslides) which acts southwards, preventing northward propagation. Each site possesses an ``instability number'' $p_{ij}$. This defines an ``inclusion probability'' $\hat{p}_{ij}:=\max(\min(p_{ij},1),0)$ which determines how easily avalanches may propagate, according to the following rules. Given that a site is the originator or ``zeroth generation'' of an avalanche, the next generation is constructed by including each east, west or south nearest neighbor site with probability equal to its respective inclusion probability. Subsequent generations are constructed by performing the inclusion test on all east, west and south nearest neighbors of the previous generation, provided they have not already been included. Sites subjected to multiple inclusion tests are included if at least one test passes. The avalanche ends when a generation has zero size. The avalanche in Figure \ref{ava} was generated by these rules.

\emph{Dynamics and the driving process.}-- We assume that each site will spontaneously originate an avalanche at rate $\hat{p}_{ij}$, so the time intervals between such events will be exponentially distributed. Avalanches take place instantaneously and the instability numbers of all involved sites are reduced by an amount $\epsilon$ immediately afterwards. This reduces the ability of the avalanche region to propagate another avalanche. We will assume that during time intervals when no avalanches take place, the instability numbers of all sites in the system follow the same random process, $\zeta(t)$, so that
\begin{equation}
dp_{ij}(t) = d\zeta(t)
\end{equation}
for all coordinates $(i,j)$. We refer to $\zeta(t)$ as the ``global driving process'', and assume that it is comprised of a combination of discontinuous upward jumps or ``shocks'', and steady increase. In the context of landslides, small shocks or steady upward drift represent background destabilizing processes like low level rainfall, snow melt and weathering \cite{High08}, whereas large but infrequent shocks represent intense rain storms, flooding and seismic activity \cite{High08}. Two important and tractable examples of jump processes which fit our assumptions are the Gamma \cite{Josh06} and compound Poisson processes \cite{Kyp06}; we will investigate the behaviour of our model using both examples.

Once $\zeta(t)$ is defined, the complete dynamics of the model may be expressed by letting $n_{ij}(t)$ be the number of times that site $(i,j)$ has been involved in an avalanche since $t=0$. We then have that:
\begin{equation}
p_{ij}(t) = \zeta(t) - \epsilon n_{ij}(t).
\end{equation}
Over time we find that the influence of the initial configuration of the instability numbers is progressively lost, and for large systems typically $p_{ij}(t) \in [0,1]$.

\emph{Simulation results.}-- We consider first the case where $\zeta(t)$ is a compound Poisson process plus a constant drift:
\begin{equation}
\label{poisson}
\zeta(t) = \nu t + \sum_{k=1}^{N(t)} J_k
\end{equation}
where $N(t)$ is a standard Poisson process with rate parameter $\lambda$, and $J_k$ is the size of the $k$th jump since $t=0$. For simplicity we will assume that jump sizes are uniformly distributed on the interval $[0,J_{max}]$. The positive constant $\nu$ is the drift rate of the process. Physically, $\nu$ determines the rate of continuous destabilisation, whereas $\lambda$ and $J_{max}$ control the frequency and magnitude of major destabilisation events.

The long term distribution of avalanche sizes is described by the probability mass function, $\psi(s)$, where $s$ is the number of sites included in an avalanche. $\psi(s)$ is estimated from simulations by recording the sizes of a large number of avalanches once the influence of the initial state has become insignificant. The examples of $\psi(s)$ in Figure \ref{COpdf} show a distinct crossover between two pure power--law scaling domains. The triangular data points correspond to the Poisson driving process. The exponent before the crossover is independent of $\zeta(t)$ whereas the crossover point, $s^\ast$, and the second exponent, labelled $\tau$, are not. We will show that both these quantities are related to the critical behaviour of the avalanche process, and to the distribution of the average value of $p_{ij}$ over the lattice, which is influenced by $\zeta(t)$.

To demonstrate that the crossover phenomenon is not unique to the Poisson driving process, we have also simulated the system in the case where $\zeta(t)$ is a Gamma process, which, conditional on $\zeta(0)=0$, has probability density:
\begin{equation}
\mathbb{P}\left[\zeta(t) \in [z,z+dz]\right] = \frac{\beta^{\alpha t} z^{\alpha t -1} e^{-\beta z}}{\Gamma(\alpha t)} dz
\end{equation}
valid for $z>0$ where the shape and rate parameters $\alpha>0$ and $\beta>0$ determine the mean $\alpha/\beta$ and variance $\alpha/\beta^2$ of the changes in $\zeta$ per unit of time. The Gamma process is a pure jump process with an infinite number of jumps in any time interval, but for any $\delta>0$ there are only a finite number larger than $\delta$ \cite{Josh06, Kyp06}. For a given mean rate of increase, a larger variance implies more variation in jump sizes.

In Figure \ref{COpdf} the simulation estimate for $\psi(s)$ using a Gamma driving process has the same exponent ($-1$) to the left of the crossover point as for the Poisson driving process. However the location of the crossover and the exponent beyond it have been altered by the properties of $\zeta(t)$. Simulation experiments show that the crossover effect is robust to the choice of jump process parameters, but can be lost if the overall driving rate or variability in jump size is too small. The parameter $\epsilon$ influences the distribution of the set $\{p_{ij}\}$ over the lattice which will develop a non--trivial correlation structure over time \cite{Dros02}. As $\epsilon \rightarrow 0$, the magnitude of local fluctuations in the instability numbers declines and therefore so does the influence of spatial correlations. However, $\epsilon$ need not be particularly small for the crossover effect to appear; it remains distinct when $\epsilon$ is increased by at least an order of magnitude compared to the cases we have considered \cite{Supp}.

\emph{Explanation of Crossover.}-- To understand the crossover, we investigate the behaviour of the spatial average instability number, $p(t) := \langle p_{ij}(t) \rangle$.
\begin{figure}
\includegraphics[width=8.6 cm]{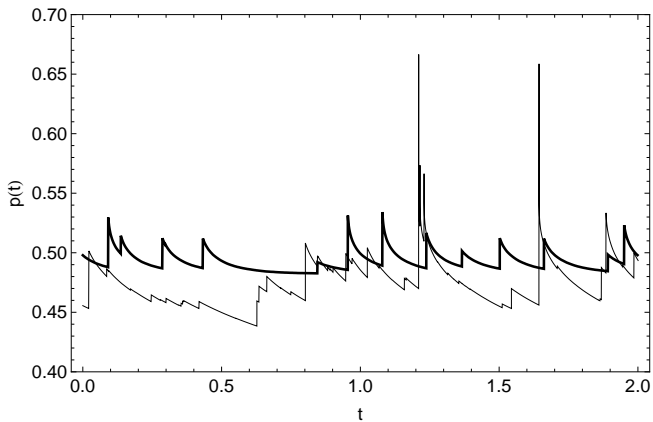}
\caption{\label{pSeries} Time series for the average instability number $p := \langle p_{ij} \rangle$ in a system of size $W \times L = 2000 \times 3000$ driven by the compound Poisson process (thick line) and Gamma process (thin line). Parameters match those used for Figure \ref{COpdf}.  }
\end{figure}
Figure \ref{pSeries} illustrates how fluctuations in $p(t)$ consist of discontinuous upward jumps due to the driving process, followed by almost continuous relaxations caused by multiple avalanches. Some particularly large upward jumps in the Gamma case are followed by almost instantaneous relaxations due to very large avalanches. This divergence in the relaxation rate is due to the existence of a critical level of average instability $p_c \approx 0.55$, beyond which, in the limit of large system size, the mean avalanche size becomes infinite. If a jump causes the system to exceed $p_c$, it will almost immediately be returned to the subcritical state.

\begin{figure}
\includegraphics[width=8.6 cm]{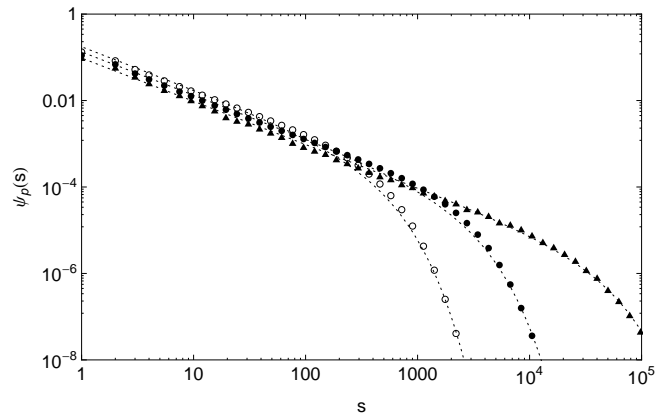}
\caption{\label{pmfs}  Estimated probability mass functions for avalanche size in a $W \times L = 2000 \times 3000$ system, sampled when $\langle p_{ij}\rangle$ lay in intervals of width $ \Delta p = 0.005$ centred about the points $ \{0.49,0.52,0.54\}$ (open circles, filled circles and triangles respectively). Functions were sampled during a simulation using the same Poisson parameters as in Figure \ref{COpdf}. The dashed lines show the scaling forms (\ref{scaling}), where parameter values were obtained by regression.  }
\end{figure}

We define $\psi_p(s)$ to be the probability mass function for avalanche size when the mean stability number is equal to $p$.  To estimate $\psi_p(s)$, the avalanche distribution was sampled when $p(t)$ lay in a series of narrow intervals and the results obtained by this method are plotted in Figure \ref{pmfs}, together with an approximate analytic form for the distribution, motivated by the following reasoning.
From Figure \ref{pmfs} we see that in common with classical branching processes and percolation \cite{Stau91}, the lattice avalanche model at given $p$ is characterised by a power--law scaling interval: $s \in [1, \xi(p)]$ where $\xi(p)$ is a cutoff size, beyond which the probability mass function decays more rapidly. Approximating this decay with an exponential we have
\begin{equation}
\label{scaling}
\psi_p(s) \approx A(p) s^{-b} \exp\left(-\frac{s}{\xi(p)}\right),
\end{equation}
where $A(p)$ is a normalising constant and $b \approx 1$ is  independent of $p$. Values for $b$ and $\xi(p)$ were determined by regression. As $\langle p_{ij} \rangle$ approaches the critical value, $p_c \approx 0.55$, the cutoff tends to infinity having approximate critical behaviour:
\begin{equation}
\label{cutoff}
\xi(p) \sim \frac{C}{(p_c-p)^\gamma} \text{ as } p \uparrow p_c,
\end{equation}
where the critical exponent $\gamma \approx 2.63 \pm 0.02$ (standard error) may be determined by linear regression on $\ln (p_c-p)$ versus $\ln \xi(p)$, and $C \approx 0.18$ is a constant. Simulation results for both Gamma and Poisson driving processes with $\epsilon = 0.01$ are consistent with this estimate.

\begin{figure}
\includegraphics[width=8.6 cm]{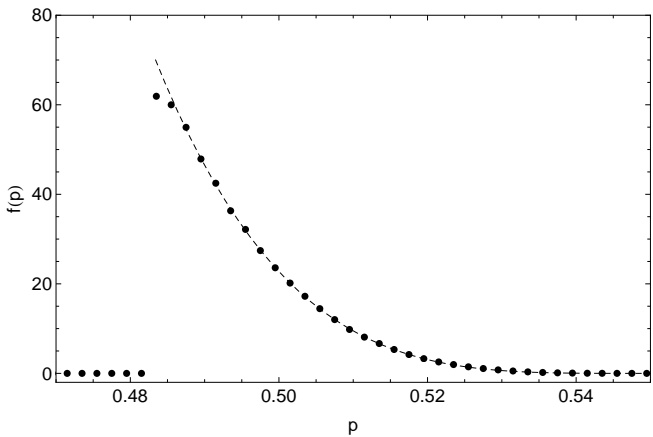}
\caption{\label{pDensity} Distribution of the average instability number $p := \langle p_{ij} \rangle$ in a system of size $W \times L = 2000 \times 3000$ driven by a compound Poisson process with parameters as in Figure \ref{COpdf}. The dashed line is the function $3.1 \times 10^6 (p_c-p)^\kappa$ where $\kappa = 3.95$ and $p_c = 0.55$. }
\end{figure}

We may now show mathematically how the crossover arises by noting that if the equilibrium probability density function of $p$ is $f(p)$, then:
\begin{equation}
\psi(s) = \int_0^\infty f(p) \psi_p(s) dp.
\end{equation}
Figure \ref{pDensity} shows a simulation estimate for $f(p)$ using the same (Poisson) driving process as in Figures \ref{COpdf} and \ref{pSeries}. Also shown is an approximate analytic form: $f(p) \propto (p_c-p)^\kappa$ for its upper tail, with $f(p>p_c)=0$, the latter condition holding due to the diverging relaxation rate for $p > p_c$. The effectiveness of this approximation near $p_c$ may be attributed to the power law divergence in the relaxation rate as $p \rightarrow p_c$. The exponent $\kappa$ must, apart from in special cases, be determined numerically.
We define $p^\ast$ to be the location of the peak of $f(p)$, finding that $p^\ast \approx 0.484$. This marks the approximate point where the constant drift component of the driving process matches the relaxation rate of the lattice.

Making use of our approximate analytic expressions for $\psi_p(s)$ and $f(p)$, and approximating $A(p)$ with a constant in the interval $[p^\ast,p_c]$, we find that
\begin{align}
\psi(s) & \propto s^{-1} \int_{p^\ast}^{p_c} (p_c-p)^\kappa \exp\left[-\frac{s (p_c-p)^\gamma}{C}\right] dp \\
& \propto s^{-\frac{1+\kappa +\gamma}{\gamma}} \int_0^{s/ \xi(p^\ast)} x^{\frac{1+\kappa-\gamma}{\gamma}} e^{-x} dx \\
&\propto \begin{cases}
s^{-1} & \text{ if } s \ll \xi(p^\ast) \\
s^{-\frac{1+\kappa+\gamma}{\gamma}} & \text{ if } s \gg \xi(p^\ast).
\label{mathCO}
\end{cases}
\end{align}
Equation (\ref{mathCO}) relates the exponents and crossover point to the properties of the underlying avalanche model and the tail exponent $\kappa$. According to this calculation, the crossover occurs at the cutoff size associated with $p^\ast$:
\begin{equation}
s^* := \xi(p^*).
\end{equation}
The power--law exponent for $s<s^\ast$ is inherited from the lattice avalanche model. The exponent for $s>s^\ast$ is:
\begin{equation}
\label{tau}
\tau := \frac{1+\kappa+\gamma}{\gamma}.
\end{equation}
For the Poisson case we have considered, equation (\ref{tau}) gives $\tau=2.9$. From Figure \ref{COpdf} we see that this exponent matches our simulation results.  The crossover point lies, theoretically, at $\xi(p^\ast) \approx 230$ which is also consistent with Figure \ref{COpdf}. We note that taking $A(p)$ to be constant amounts to ignoring, for $s>s^\ast$, a correction of magnitude $(\ln s)^{-1}$ to the gradient of $\psi(s)$ on a log-log graph. In the case of a Gamma driving process, $f(p)$ possesses the same form of upper tail behaviour, so the exponent $\kappa$ may be found, and equation (\ref{tau}) correctly gives the large $s$ exponent (see Figure \ref{COpdf}). For smaller $p$, in contrast to the Poisson case, $f(p)$ is approximately Gaussian, however we find that our expression for $s^\ast$ accurately predicts the crossover location.

\emph{Discussion and Conclusion.}-- We have introduced a model of spatial failure avalanches where the failure probability is a local dynamical variable, driven upwards by a global random process, and declining locally where avalanches occur. Shocks in the driving process periodically throw the system into a very unstable condition, from which it quickly retreats due to large failure events, leaving behind a power--law crossover signature. The crossover point, $s^\ast$, and second exponent, $\tau$, are determined by both the critical behaviour of the system and the characteristics of the driving process. Broadly speaking, the exponent $\tau$ is reduced by a greater frequency of larger shocks in the driving process, whereas the crossover point reflects the long term, lower level driving rate (see supplementary material \cite{Supp} for more detail).  A crossover observed in real data, for example in landslide inventories or epidemic records, might therefore provide insight into the frequency with which such systems were subject to major destabilising influences, and their typical proximity to criticality.

We may place our avalanche rules in the context of previous work by noting that they bear a similarity to a number of models \cite{Bak88,Olam92,Herg03,Herg12_2} including diffusion  percolation (DP) (closely related to bootstrap percolation) \cite{Adler88,Adler91,Aize88,Chav95}. The analogy to percolation is made by viewing our $\{\hat{p}_{ij}\}$ as initial occupation probabilities of the lattice. In DP, additional sites are occupied if they have $k$ or more occupied neighbors, mimicking multiple inclusion tests in our rules. The analogue of our avalanche construction on a lattice whose occupation state was already determined would be to select an occupied site, and determine the size of the cluster to which it belonged. For DP this process would produce numbers equivalent to our $b$ and $\gamma$, (see equations (\ref{scaling}) and (\ref{cutoff})), of $1.05$ and $2.53$ respectively, universal for percolation \cite{Chav95,Stau91}. 

In previous models of landslides, the importance of the rate at which the system is driven has been recognised \cite{Pie06_2,Pieg06,Herg00,Herg13,Herg13}, and shown to produce a transition from power--law to non power--law behavior as it is increased \cite{Pieg06,Pie06_2} and to capture ``roll-over'' deviations \cite{Mala04} at small event sizes \cite{Pieg06}. In cascading models of landslides and other natural hazards, a crossover from power-law to exponential or other non power--law decay at a particular event size \cite{Olam92,Turc02,Herg12_2} is observed due to system size effects. The new effect that we have demonstrated is a crossover from one power--law to another. This occurs due to the interaction between dramatic driving events and the near critical behaviour of the system, which controls the second power--law exponent, $\tau$, via the cutoff exponent $\gamma$.



\end{document}